\documentclass[11pt]{article}
\usepackage{enumerate}
\usepackage[shortlabels]{enumitem}
\usepackage{fullpage}
\usepackage{amsmath,amsthm,amssymb}
\usepackage{enumerate}
\usepackage{rotating}
\usepackage[colorlinks=true,
            linkcolor=red,
            urlcolor=green,
            citecolor=blue]{hyperref}

\usepackage{cancel}
\usepackage{framed}
\usepackage{algorithm}
\usepackage[noend]{algorithmic}
\usepackage{graphicx}
\usepackage{subfigure}
\usepackage{float}

\newcommand{\cF}{{\cal F}}
\newcommand{\cD}{{\cal D}}
\newcommand{\cO}{{\cal O}}
\newcommand{\cC}{{\cal C}}

\newtheorem{theorem}{Theorem}

\newcommand{\eps}{\varepsilon}

\ifdefined\DEBUG
\newcommand{\fab}[1]{\textcolor{red}{#1}}
\else
  \newcommand{\fab}[1]{#1}
\fi

\title{A Refined Approximation for\\Euclidean k-Means}
\author{Fabrizio Grandoni\footnote{IDSIA, USI-SUPSI} \and Rafail Ostrovsky\footnote{UCLA} \and Yuval Rabani\footnote{The Hebrew University of Jerusalem} \and Leonard J. Schulman\footnote{Caltech} \and Rakesh Venkat\footnote{IIT Hyderabad}}

\date{\empty}

\begin{document}
\maketitle

\begin{abstract}
\noindent In the Euclidean $k$-Means problem we are given a collection of $n$ points $\cD$ in an Euclidean space and a positive integer $k$. Our goal is to identify a collection of $k$ points in the same space (centers) so as to minimize the sum of the squared Euclidean distances between each point in $\cD$ and the closest center. This problem is known to be APX-hard and the current best approximation ratio is a primal-dual $6.357$ approximation based on a standard LP for the problem [Ahmadian et al. FOCS'17, SICOMP'20].

In this note we show how a minor modification of Ahmadian et al.'s analysis leads to a slightly improved $6.12903$ approximation. As a related result, we also show that the mentioned LP has integrality gap at least $\frac{16+\sqrt{5}}{15}>1.2157$.
\end{abstract}

\section{Introduction}
\label{sec:intro}

%\subsection{Euclidean $k$-Means} \label{sec:ekm}

Clustering is a central problem in Computer Science, with many applications in data science, machine learning etc. One of the most famous and best-studied problems in this area is Euclidean $k$-Means: given a set $\cD$ of $n$ points (or \emph{demands}) in $\mathbb{R}^{\ell}$ and an integer $k\in [1,n]$, select $k$ points $S$ (\emph{centers}) so as to minimize $\sum_{j\in \cD}d^2(j,S)$. Here $d(j,i)$ is the Euclidean distance between points $j$ and $i$ and for a set of points $I$, $d(j,I)=\min_{i\in I}d(j,i)$.
In other words, we wish to select $k$ centers so as to minimize the sum of the squared Euclidean distances between each demand and the closest center. Equivalently, a feasible solution is given by a partition of the demands into $k$ subsets (\emph{clusters}). The cost $w_C$ of a cluster $C\subset \cD$ is $\sum_{j\in C}d^2(c,\mu)$, where $\mu$ is the center of mass of $C$. We recall that $w_C$ can also be expressed as $\frac{1}{2|C|}\sum_{j\in C}\sum_{j'\in C}d^2(j,j')$. Our goal is to minimize the total cost of these clusters.

Euclidean $k$-Means is well-studied in terms of approximation algorithms. It is known to be APX-hard. More precisely, it is hard to approximate $k$-Means below a factor $1.0013$ in polynomial time unless $P=NP$ \cite{ACKS15,LSW17}. The hardness was improved to $1.07$ under the Unique Games Conjecture \cite{CK19}. Some heuristics are known to perform very well in practice, however their approximation factor is $O(\log k)$ or worse on general instances \cite{AV06,AV07,L82,V11}. Constant approximation algorithms are known. A local-search algorithm by Kanugo et al.~\cite{KMNPSW04} provides a $9+\eps$ approximation\footnote{Throughout this paper by $\eps$ we mean an arbitrarily small positive constant. W.l.o.g.\ we assume $\eps\leq 1$.}. The authors also show that natural local-search based algorithms cannot perform better than this. This ratio was improved to $6.357$ by Ahmadian et al.~\cite{ANSW17,ANSW20} using a primal-dual approach. They also prove a $9+\eps$ approximation for general (possibly non-Euclidean) metrics. Better approximation factors are known under reasonable restrictions on the input \cite{ABS10,BBG09,CKM19,ORSS12}. A PTAS is known for constant $k$ \cite{M00} or for constant dimension $\ell$ \cite{CKM19,FRS19}. Notice that $\ell$ can be always assumed to be $O(\log n)$ by a standard application of the Johnson-Lindenstrauss transform \cite{JL84}. This was recently improved to $O(\log k+\log\log n)$ \cite{BBCGS19} and finally to $O(\log k)$ \cite{MMR19}.

In this paper we describe a simple modification of the analysis of Ahmadian et al.~\cite{ANSW20} which leads to a slightly improved approximation for Euclidean $k$-Means (see Section \ref{sec:improvedApx}).
\begin{theorem}\label{thr:apxMain}
There exists a deterministic polynomial-time algorithm for Euclidean $k$-Means with approximation ratio $\rho+\eps$ for any positive constant $\eps>0$, where $$\rho:=\left(1+\sqrt{\frac{1}{2}(2+\sqrt[3]{3-2\sqrt{2}}+\sqrt[3]{3+2\sqrt{2}})}\right)^2<6.12903.$$
\end{theorem}

The above approximation ratio is w.r.t.\ the optimal fractional solution to a standard LP relaxation $LP_{\text{k-Means}}$ for the problem (defined later). As a side result (see Section \ref{sec:lb}), we prove a lower bound on the integrality gap of this relaxation (we are not aware of any explicit such lower bound in the literature).
%\fabr{Following Leonard's concerns, I removed "Our lower bound (or better, the fact that we did not manage to prove a larger lower bound) might be seen as a positive evidence that much better approximation algorithms based on the same LP might exist."}
%Our lower bound (or better, the fact that we did not manage to prove a larger lower bound) might be seen as a positive evidence that much better approximation algorithms bases on the same LP might exist.
\begin{theorem}\label{thr:LB}
The integrality gap of $LP_{\text{k-Means}}$, even in the Euclidean plane (i.e., for $\ell=2$), is at least $\frac{16+\sqrt{5}}{15}>1.2157$.
\end{theorem}

\subsection{Preliminaries}

%Since our approach builds heavily on \cite{ANSW20}, we will try to use a notation as close as possible to the latter paper.
As mentioned earlier, one can formulate Euclidean $k$-Means in term of the selection of $k$ centers. In this case, it is convenient to discretize the possible choices for the centers, hence obtaining a polynomial-size set $\cF$ of candidate centers, at the cost of an extra factor $1+\eps$ in the approximation ratio (we will neglect this factor in the approximation ratios since it is absorbed by analogous factors in the rest of the analysis). In particular we will use the construction in \cite{DKKR03} (Lemma 24) that chooses as $\cF$ the centers of mass of any collection of up to $16/\eps^2$ points with repetitions. In particular $|\cF|=O(n^{16/\eps^2})$ in this case.

%The approximation algorithm that we will describe works with no relevant changes in the variant of Euclidean  case where the centers need to be points in $\cD$ (in this case, simply set $\cF=\cD$).

Let $c(j,i)$ be an abbreviation for $d^2(j,i)$. Then a standard LP-relaxation for $k$-Means is as follows:
\begin{align*}
\min & \sum_{i\in \cF,j\in \cD}x_{ij}\cdot c(j,i)& LP_{\text{k-Means}} \\
s.t. & \sum_{i\in \cF}x_{ij}\geq 1 & \forall j\in \cD \\
      & x_{ij}\leq y_i & \forall j\in \cD, \forall i\in \cF \\
      & \sum_{i\in \cF}y_i\leq k & \forall j\in \cD, \forall i\in \cF \\
      & x_{ij},y_i\geq 0 & \forall j\in \cD, \forall i\in \cF
\end{align*}
In an integral solution, we interpret $y_i=1$ as $i$ being a selected center in $S$ ($i$ is \emph{open}), and $x_{ij}=1$ as demand $j$ being \emph{assigned} to center $i$\footnote{Technically each demand is automatically assigned to the closest open center. However it is convenient to allow also sub-optimal assignments in the LP relaxation.}. The first family of constraints states that each demand has to be assigned to some center, the second one that a demand can only be assigned to an open center, and the third one that we can open at most $k$ centers.

For any parameter $\lambda>0$ (\emph{Lagrangian multiplier}), the \emph{Lagrangian relaxation} $LP(\lambda)$ of $LP_{\text{k-Means}}$ (w.r.t. the last matrix constraint) and its dual $DP(\lambda)$ are as follows:
\begin{align*}
\min & \sum_{i\in \cF,j\in \cD}x_{ij}\cdot c(j,i)+\lambda\cdot \sum_{i\in \cF}y_i - \lambda\,k & LP(\lambda) \\
s.t. & \sum_{i\in \cF}x_{ij}\geq 1 & \forall j\in \cD \\
      & x_{ij}\leq y_i & \forall j\in \cD, \forall i\in \cF \\
      & x_{ij},y_i\geq 0 & \forall j\in \cD, \forall i\in \cF
\end{align*}
\begin{align}
\max & \sum_{j\in \cD}\alpha_{j} -\lambda k& DP(\lambda) \nonumber\\
s.t. & \sum_{j\in \cD}\max\{0,\alpha_j-c(j,i)\}\leq \lambda & \forall i\in \cF \label{dual:1}\\
  & \alpha_{j}\geq 0 & \forall j\in \cD \nonumber
\end{align}
Above $\max\{0,\alpha_j-c(j,i)\}$ replaces the dual variable $\beta_{ij}$ corresponding to the second constraint in the primal in the standard formulation of the dual LP. Notice that, by removing the fixed term $-\lambda k$ in the objective functions of $LP(\lambda)$ and $DP(\lambda)$, one obtains the standard LP relaxation $LP_{FL}(\lambda)$ for the Facility Location problem (FL) with uniform facility cost $\lambda$ and its dual $DP_{FL}(\lambda)$.

We say that a $\rho$-approximation algorithm for a FL instance of the above type is \emph{Lagrangian Multiplier Preserving} (LMP) if it returns a set of facilities $S$ that satisfies:
$$
\sum_{j\in \cD}c(j,S) \leq \rho(OPT(\lambda)-\lambda |S|),
$$
where $OPT(\lambda)$ is the value of the optimal solution to $LP_{FL}(\lambda)$.

%\section{An Improved LMP Approximation for Euclidean Facility Location}

\section{A Refined Approximation for Euclidean $k$-Means}
\label{sec:improvedApx}

In this section we present our refined approximation for Euclidean $k$-Means. We start by presenting the LMP approximation algorithm for the FL instances arising from $k$-Means described in \cite{ANSW20} in Section \ref{sec:algorithm}. We then present the analysis of that algorithm as in \cite{ANSW20} in Section \ref{sec:oldAnalysis}. In Section \ref{sec:newAnalysis} we describe our refined analysis of the same algorithm. Finally, in Section \ref{sec:kMeans} we sketch how to use this to approximate $k$-Means.

\subsection{A Primal-Dual LMP Algorithm for Euclidean Facility Location}
\label{sec:algorithm}

We consider an instance of Euclidean FL induced by a $k$-Means instance in the mentioned way, for a given Lagrangian multiplier $\lambda>0$.

%We consider the following primal and dual LPs:\fabr{ANSW use this notation for the Lagrangian relaxation}
%\begin{align*}
%\min & \sum_{i\in \cF,j\in \cD}x_{ij}\cdot c(j,i)+\lambda\cdot \sum_{i\in \cF}y_i & LP(\lambda) \\
%s.t. & \sum_{i\in \cF}x_{ij}\geq 1 & \forall j\in \cD \\
%      & x_{ij}\leq y_i & \forall j\in \cD, \forall i\in \cF \\
%      & x_{ij},y_i\geq 0 & \forall j\in \cD, \forall i\in \cF
%\end{align*}
%\begin{align}
%\max & \sum_{j\in \cD}\alpha_{j} & DUAL(\lambda) \nonumber\\
%s.t. & \sum_{j\in \cD}\max\{0,\alpha_j-c(j,i)\}\leq \lambda & \forall i\in \cF \label{dual:1}\\
%  & \alpha_{j}\geq 0 & \forall j\in \cD \nonumber
%\end{align}
%Above $\max\{0,\alpha_j-c(j,i)\}$ replaces the dual variable $\beta_{ij}$ in the standard formulation of the dual LP.

We consider exactly the same Lagrangian Multiplier Preserving (LMP) primal-dual algorithm $JV(\delta)$ as in \cite{ANSW20}. In more detail, let $\delta\geq 2$ be a parameter to be fixed later. The algorithm consists of a dual-growth phase and a pruning phase. The dual-growth phase is exactly as in the classical primal-dual algorithm JV by Jain and Vazirani \cite{JV01}. We start with all the dual variables set to $0$ and an empty set $\cO_t$ of tentatively open facilities. The clients such that $\alpha_j\geq c(j,i)$ for some $i\in F'$ are frozen, and the other clients are active. We grow the dual variables of active clients at uniform rate until one of the following two events happens. The first event is that some constraint of type \eqref{dual:1} becomes tight. At that point the corresponding facility $i$ is added to $\cO_t$ and all clients $j$ with $\alpha_j\geq c(j,i)$ are set to frozen. The second event is that $\alpha_j=c(j,i)$ for some some $i\in \cO_t$. In that case $j$ is set to frozen. In any case, the facility $w(j)$ that causes $j$ to become frozen is called the witness of $j$. The phase halts when all clients are frozen.

In the pruning phase we will close some facilities in $\cO_t$, hence obtaining the final set of open facilities $IS$. Here $JV(\delta)$ deviates from $JV$. For each client $j\in \cD$, let $N(j)=\{i\in \cF: \alpha_j>c(j,i)\}$ be the set of facilities $i$ such that $j$ contributed with a positive amount to the opening of $i$. Symmetrically, for $i\in \cF$, let $N(i)=\{j\in \cD: \alpha_j>c(j,i)\}$ be the clients that contributed with a positive amount to the opening of $i$. % \ljs{edited, this said $j$}.
For $i\in \cO_t$, we let $t_i=\max_{j\in N(i)}\alpha_j$, where the values $\alpha_j$ are considered at the end of the dual-growth phase. We set conventionally $t_i=0$ for $N(i)=\emptyset$. Intuitively, $t_i$ is the ``time'' when facility $i$ is tentatively open (at which point all the dual variables of contributing clients stop growing). We define a conflict graph $H$ over tentatively open facilities as follows. The node set of $H$ is $\cO_t$. We place an edge between $i,i'\in \cO_t$ iff the following two conditions hold: (1) for some client $j$, $j\in N(i)\cap N(i')$ (in words, $j$ contributes to the opening of both $i$ and $i'$) and (2) one has $c(i,i')\leq \delta\cdot \min\{t_i,t_{i'}\}$. In this graph we compute a maximal independent set $IS$, which provides the desired solution to the facility location problem (where each client is assigned to the closest facility in $IS$).

We remark that the pruning phase of $JV$ differs from the one of $JV(\delta)$ only in the definition of $H$, where condition (2) is not required to hold (or, equivalently, $JV$ behaves like $JV(+\infty)$ for $\lambda>0$).

%We next describe the analysis in [ANSW], and then how we improve on it. We start with the more complex k-Means case, and then move to k-Median,

\subsection{The Analysis in \cite{ANSW20}}
\label{sec:oldAnalysis}

%We describe the analysis in [ANSW] to highlight the differences with respect to our analysis.
The general goal is to show that
$$
\sum_{j\in \cD}c(j,IS)\leq \rho(\sum_{j\in \cD}\alpha_j-\lambda |IS|),
$$
for some $\rho\geq 1$ as small as possible. This shows that the algorithm is an LMP $\rho$-approximation for the problem. It is sufficient to prove that, for each client $j$, one has
$$
\frac{c(j,IS)}{\rho}\leq \alpha_j-\sum_{i\in N(j)\cap IS}(\alpha_j-c(j,i))=\alpha_j-\sum_{i\in IS}\max\{0,\alpha_j-c(j,i)\}.
$$
Let $S=N(j)\cap IS$ and $s=|S|$. We distinguish $3$ cases depending on the value of $s$:

\paragraph{Case A: $s=1$.} Let $S=\{i^*\}$. Then for any $\rho\geq 1$,
$$
\frac{c(j,IS)}{\rho}\leq c(j,IS)=c(j,i^*)=\alpha_j-(\alpha_j-c(j,i^*)).
$$

\paragraph{Case B: $s>1$.} Here we use the properties of Euclidean metrics. The sum $\sum_{i\in S}c(j,i)$ is the sum of the squared distances from $j$ to the facilities in $S$. This quantity is lower bounded by the sum of the squared distances from $S$ to the centroid $\mu$ of $S$. Recall that $\sum_{i\in S}c(\mu,i)=\frac{1}{2s}\sum_{i\in S}\sum_{i'\in S}c(i,i')$. We also observe that, by construction, for any two distinct $i,i'\in IS$ one has
$$
c(i,i')>\delta \cdot \min \{t_i,t_{i'}\}\geq \delta \cdot \alpha_j,
$$
where the last inequality follows from the fact that $j$ is contributing to the opening of both $i$ and $i'$. Altogether one obtains
$$
\sum_{i\in S}c(j,i)\geq \sum_{i\in S}c(\mu,i)=\frac{1}{2s}\sum_{i\in S}\sum_{i'\in S}c(i,i')\geq \frac{(s-1)\delta \alpha_j}{2}.
$$
Thus
$$
\sum_{i\in S}(\alpha_j-c(j,i))\leq (s-\frac{\delta(s-1)}{2})\alpha_j=(s(1-\frac{\delta}{2})+\frac{\delta}{2})\alpha_j \overset{\delta\geq 2,s\geq 2}\leq (2-\frac{\delta}{2})\alpha_j.
$$
Using the fact that $\alpha_j> c(j,i)$ for all $i\in S$, hence $\alpha_j>c(j,IS)$, one gets
$$
(\frac{\delta}{2}-1)c(j,IS)\overset{\fab{\delta\geq 2}}{\leq} (\frac{\delta}{2}-1)\alpha_j.
$$
We conclude that
$$
\sum_{i\in S}(\alpha_j-c(j,i))+(\frac{\delta}{2}-1)c(j,IS)\leq (2-\frac{\delta}{2})\alpha_j+(\frac{\delta}{2}-1)\alpha_j=\alpha_j.
$$
This gives the desired inequality assuming that $\rho\geq \frac{1}{\delta/2-1}$.

\paragraph{Case C: $s=0$.} Consider the witness $i=w(j)$ of $j$. Notice that $\alpha_j\geq t_{i}$ and $\alpha_j\geq c(j,i)=d^2(j,i)$. Hence
$$
d(j,i)+\sqrt{\delta t_{i}} \leq (1+\sqrt{\delta})\sqrt{\alpha_j}.
$$
If $i\in IS$, then $d(j,IS)\geq d(j,i)$. Otherwise there exists $i'\in IS$ such that $d^2(i,i')\leq \delta \min\{t_i,t_{i'}\} \leq \delta t_i$. Thus
$
d(j,IS)\leq d(j,i)+d(i,i')\leq d(j,i)+\sqrt{\delta t_i}.
$
In both cases one has
$
d(j,IS)\leq (1+\sqrt{\delta})\sqrt{\alpha_j},
$
hence
$$
c(j,IS)\leq (1+\sqrt{\delta})^2\alpha_j.
$$
This  gives the desired inequality for $\rho\geq (1+\sqrt{\delta})^2$.

\paragraph{Fixing $\delta$.} Altogether we can set $\rho=\max\{\frac{1}{\delta/2-1},(1+\sqrt{\delta})^2\}$. The best choice for $\delta$ (namely, the one that minimizes $\rho$) is the solution of $\frac{1}{\delta/2-1}=(1+\sqrt{\delta})^2$. This is achieved for $\delta\simeq 2.3146$ and gives $\rho\simeq 6.3574$.

\subsection{A Refined Analysis}
\label{sec:newAnalysis}

We refine the analysis in Case B as follows. Let $\Delta=\sum_{i\in S}c(j,i)$. \fab{We already proved that, for $\delta\geq 2$, $\alpha_j-\sum_{i\in S}(\alpha_j-c(j,i))\geq 0$}. \fab{Hence it is sufficient} to upper bound
$$
\frac{c(j,S)}{\alpha_j-\sum_{i\in S}(\alpha_j-c(j,i))}=\frac{c(j,S)}{\sum_{i\in S}c(j,i)-(s-1)\alpha_j}=\frac{c(j,S)}{\Delta-(s-1)\alpha_j}.
$$
Instead of using the upper bound $c(j,S)\leq \alpha_j$ we use the average
$$
c(j,S)\leq \frac{1}{s}\sum_{i\in S}c(j,i)=\frac{\Delta}{s}.
$$
Then it is sufficient to upper bound
$$
\frac{1}{s}\frac{\Delta}{\Delta-(s-1)\alpha_j}.
$$
The derivative in $\Delta$ of the above function is $\frac{1}{s}\frac{-(s-1)\alpha_j}{(\Delta-(s-1)\alpha_j)^2}<0$. Hence the maximum is achieved for the smallest possible value of $\Delta$. Recall that we already showed that $\Delta\geq \frac{(s-1)\delta \alpha_j}{2}$. Hence a valid upper bound is
$$
\frac{1}{s}\frac{\frac{(s-1)\delta \alpha_j}{2}}{\frac{(s-1)\delta \alpha_j}{2}-(s-1)\alpha_j}=\frac{1}{s}\frac{\delta/2}{\delta/2-1}\overset{s\geq 2}{\leq}\frac{\delta/4}{\delta/2-1}.
$$
This imposes $\rho\geq \frac{\delta/4}{\delta/2-1}$ rather than $\rho\geq \frac{1}{\delta/2-1}$ in Case B. Notice that this is an improvement for $\delta<4$. The best choice of $\delta$ is now obtained by imposing $\frac{\delta/4}{\delta/2-1}=(1+\sqrt{\delta})^2$. This gives $\delta=\frac{1}{2}(2+\sqrt[3]{3-2\sqrt{2}}+\sqrt[3]{3+2\sqrt{2}})\simeq 2.1777$ and $\rho=\left(1+\sqrt{\frac{1}{2}(2+\sqrt[3]{3-2\sqrt{2}}+\sqrt[3]{3+2\sqrt{2}})}\right)^2<6.12903$.

\subsection{From Facility Location to k-Means}
\label{sec:kMeans}

We can use the refined $\rho:=\left(1+\sqrt{\frac{1}{2}(2+\sqrt[3]{3-2\sqrt{2}}+\sqrt[3]{3+2\sqrt{2}})}\right)^2$ approximation for Euclidean Facility Location from previous section to derive a $\rho+\eps$ approximation for Euclidean k-Means, for any constant $\eps>0$. Here we follow the approach of \cite{ANSW20} with only minor changes. In more detail, the authors consider a variant of the FL algorithm described before, whose approximation factor is $\rho+\eps$ rather than $\rho$. A careful use of this algorithm leads to a solution opening precisely $k$ facilities, which leads to the desired approximation factor. In their analysis the authors use slight modifications of the inequality $(1+\sqrt{\delta})\sqrt{\alpha_j}\geq d(j,i)+\sqrt{\delta t_i}$ (coming from Case C, which is the same in their and our analysis). The goal is to prove that the modified algorithm is $\rho+\eps$ approximate. Here $\delta$ and $\rho$ are used as parameters. Therefore it is sufficient to replace their values of these parameters with the ones coming from our refined analysis. The rest is identical.

\section{Lower Bound on the Integrality Gap}
\label{sec:lb}

In this section we describe our lower bound instance for the integrality gap of $LP_{\text{k-Means}}$. It is convenient to consider first the following slightly different relaxation, based on clusters (with $w_C$ as defined in Section \ref{sec:intro}):
%The above approximation ratio is w.r.t. to the optimal fractional solution to the following natural LP relaxation $LP'_{\text{k-Means}}$ for the problem\footnote{As discussed later, we use a slightly different relaxation which is however equivalent up to $1+\eps$ factors.}:
\begin{align*}
\min & \sum_{C\in \cC}w_Cx_{C}& LP'_{\text{k-Means}} \\
s.t. & \sum_{C\in \cC:j\in C}x_{C}\geq 1 & \forall j\in \cD \\
      & \sum_{C\in \cC}x_C\leq k \\
      & x_{C}\geq 0 & \forall C\in \cC
\end{align*}
Here $\cC$ denotes the set of possible clusters, i.e. the possible subsets of points. In an integral solution $x_C=1$ means that cluster $C$ is part of our solution.
%Though $LP'_{\text{k-Means}}$ has an exponential number of variables it can be solved exactly in polynomial time.\fabr{Exactly or approximately?}
%We remark that this is larger than the known hardness of approximation results for this problem \fab{\cite{???}}).\fabr{New Vincent's results}

Our instance is on the Euclidean plane, and its points are the (10) vertices of two regular pentagons of side length $1$. These pentagons are placed so that any two vertices of distinct pentagons are at large enough distance $M$ to be fixed later. Here $k=5$. We remark that our argument can be easily extended to an arbitrary number of points by taking $2h$ such pentagons for any integer $h\geq 1$ so that the pairwise distance between vertices of distinct pentagons is at least $M$, and setting $k=5h$.

A feasible fractional solution is obtained by setting $x_C=0.5$ for every $C$ consisting of a pair of consecutive vertices in the same pentagon (so we are considering $10$ fractional clusters in total). Obviously this solution is feasible. The cost $w_C$ of each such cluster $C$ is $2(0.5)^2=0.5$. Hence the cost of this fractional solution is $10\cdot 0.5\cdot 0.5=\frac{5}{2}$.

Next consider the optimal integral solution, consisting of $5$ clusters. Recall that the radius of each pentagon (i.e. the distance from a vertex to its center) is $r=\sqrt{\frac{2}{5-\sqrt{5}}}\simeq 0.851$ and the distance between two non-consecutive vertices in the same pentagon is $d=\frac{\sqrt{5}+1}{2}\simeq 1.618$. A solution with two clusters consisting of the vertices of each pentagon costs $10r^2$. Any cluster involving vertices of distinct pentagons costs at least $M^2/2$, hence for $M$ large enough the optimal solution forms clusters only with vertices of the same pentagon. In more detail the optimal solution consists of $x\in \{1,2,3,4\}$ clusters containing the vertices of one pentagon and $5-x$ clusters containing the vertices of the remaining pentagon. Let $w(x)$ be the minimum cost associated with one pentagon assuming that we form $x$ clusters with its vertices. Clearly $w(1)=5r^2=\frac{10}{5-\sqrt{5}}$. Regarding $w(4)$, it is obviously convenient to choose two consecutive vertices in the unique cluster of size $2$. Thus $w(4)=1/2$.
%$\frac{3(1+d^2)}{4}=\frac{15+3\sqrt{5}}{8}$.
%\fabr{w(4) comes from 4 (necessarily consecutive) vertices of one polygon. I summed up the squared distances (1 or $d^2$) and divided by 4}
For $x\in \{2,3\}$,
% it is always convenient to form
we note, as is easy to verify, that
clusters with consecutive vertices are less expensive than the alternatives. %\ljs{slight rewrite}
For $w(2)$, one might form one cluster of size $1$ and one of size $4$. This would cost $\frac{3(1+d^2)}{4}=\frac{15+3\sqrt{5}}{8}$. Alternatively, one might form one cluster of size $2$ and one of size $3$, at smaller cost $\frac{1}{2}+\frac{2+d^2}{3}=\frac{10+\sqrt{5}}{6}$. Thus $w(2)=\frac{10+\sqrt{5}}{6}$. For $x=3$, one might form two clusters of size $1$ and one of size $3$, or two clusters of size $2$ and one of size $1$. The associated cost in the two cases is $\frac{2+d^2}{3}>1$ and $2\cdot\frac{1}{2}=1$, resp. Hence $w(3)=1$. So the overall cost of the optimal integral solution is $\min\{w(1)+w(4),w(2)+w(3)\}=w(2)+w(3)=\frac{16+\sqrt{5}}{6}$. Thus the integrality gap of $LP'_{\text{k-Means}}$ is at least $\frac{16+\sqrt{5}}{6}\cdot \frac{2}{5}=\frac{16+\sqrt{5}}{15}$.

Consider next $LP_{\text{k-Means}}$. Here a technical complication comes from the definition of $\cF$ which is not part of the input instance of $k$-Means. The same construction as above works if we let $\cF$ contain the centers of mass of any set of $2$ or $3$ points. Notice that this is automatically guaranteed by the construction in \cite{DKKR03} for $16/\eps^2\geq 3$. In this case the optimal integral solutions to $LP_{\text{k-Means}}$ and $LP'_{\text{k-Means}}$ are the same in the considered example. Furthermore one obtains a feasible fractional solution to $LP_{\text{k-Means}}$ of cost $5/2$ by setting $y_i=0.5$ for the centers of mass of any two consecutive vertices of the same pentagon, and setting $x_{ij}=0.5$ for each point $i$ and the two closest centers $j$ with positive $y_j$. This concludes the proof of Theorem \ref{thr:LB}.

\section*{Acknowledgments}
Work supported in part by the NSF grant 1909972 and the SNF Excellence Grant 200020B\_182865/1.

\bibliographystyle{plain}
\bibliography{kMeansBib}

\end{document}